\documentclass[fleqn,usenatbib]{mnras}

\usepackage{newtxtext,newtxmath}

\usepackage[T1]{fontenc}
\usepackage{ae,aecompl}

\usepackage{graphicx}   

\usepackage{color}
\usepackage{bm}

\newcommand\be{\begin{equation}}
\newcommand\en{\end{equation}}
\newcommand\msun{M_{\odot}}

\newcommand\lsun{L_{\odot}}
\newcommand\mdot{\dot{M}}
\newcommand\msunyr{M_{\odot}\, {\rm yr^{-1}}}

\newcommand\twco{$\rm ^{12}CO$}
\newcommand\tbol{T$_{\rm bol}$}
\newcommand\lbol{L$_{\rm bol}$}

\usepackage[dvipsnames]{xcolor}

\title[protostar mass-luminosity] {On the protostellar mass-luminosity relation}
  \author[Hartmann et al.]
      {Lee Hartmann$^{1}$, John J. Tobin$^{2}$, Patrick Sheehan$^{2}$,
      Marina Kounkel$^{3}$,
      Claire Zhao$^4$
\\
$^{1}$ Department of Astronomy, University of Michigan, 1085 S. University Ave., Ann Arbor, MI 48109, USA \\
$^{2}$ 
National Radio Astronomy Observatory, 520 Edgemont Rd., Charlottesville,VA 22903, USA \\
$^3$ Department of Physics and Astronomy, University of North Florida, 1 UNF Dr, Jacksonville, FL, 32224, USA\\
$^4$ Lakeside School, Seattle, WA, USA
}
    
\date{Accepted XXX. Received YYY; in original form ZZZ}

\pubyear{2022}

\begin{document}

\label{firstpage}
\pagerange{\pageref{firstpage}--\pageref{lastpage}} \maketitle

\begin{abstract}
We present a preliminary view of the protostellar mass-luminosity relation using current samples of protostars with dynamical mass estimates.
To provide a lower limit to the expected luminosities, we adopt
an empirical estimate for the
intrinsic (without accretion) protostellar luminosity and radius as a function of mass. 
We find that many of the protostars with current dynamical mass estimates track the empirical mass-luminosity "birthline" reasonably closely, suggesting that their accretion luminosities may be at most comparable to their photospheric
radiation.
In turn, this implies that mass accretion rates for many objects are well below that required to build up the final stellar mass in typical estimated protostellar lifetimes. 
A small subset of the protostars have luminosities well above the predicted photospheric values, consistent with evolutionarily-important mass addition.
These results hint at a possible bimodal distribution of accretion, but a firm conclusion is not possible given the small size of and likely biases in the current sample.
\end{abstract}


\section{Introduction}
\label{sec:intro}

The most fundamental parameter of a protostar is its mass. Direct
measurements of protostar masses $M_*$ to compare with
the mass contained in their remnant
envelopes are essential to establish their evolutionary states \citep[e.g.,][]{andre93,sheehan22}. Moreover, protostar masses, 
in combination with system luminosities
$L_{bol}$, are needed to
estimate mass accretion rates $\mdot$ to improve our
understanding of stellar mass buildup, via the
relation
\begin{equation}
L_{bol} = L_* + L_{acc} 
= L_* + \eta G M_* \mdot /R_* \,.
\end{equation}
Here the protostellar 
photospheric luminosity  $L_*$ and the
radius $R_*$ can be estimated from either
theoretical or empirical mass-luminosity and mass-radius relations for young stars, assuming a
suitable factor $\eta < 1$ for the efficiency
of converting accretion energy into radiation.
Determining accretion rates for individual
objects as a function of mass
is essential to assessing whether the 
observed distributions of protostellar
luminosities imply that much of 
the mass accretion occurs in short, strong
bursts \citep[e.g.,][]{kenyon90,kenyon94, vorobyov06,vorobyov10}, or can be explained
by smoothly-varying (though non-constant)
accretion \citep{offner11,myers14,fischer17}; \citep[see][for a recent review]{fischer23}.

Sensitive, high-spatial resolution interferometry offers a means to address these problems by determining
dynamical masses from the Keplerian 
rotation of a spatially-resolved protostellar disks.
Substantial progress has been
made in determining masses for low-mass protostars \citep[e.g.,][]{lommen08,takakuwa12,murillo13,ohashi14,harsono14,chou14,
chou16,brinch16,yen17,lee17,tobin20b,maureira20,reynolds21,cheng22,vanthoff23,yamato23,kido23,sai23,
sharma23,aso23,thieme23,
flores23,
santamaria23,
han23},
mostly coming from the ALMA eDisk large program \citep{ohashi23}.
Generally, observations of \twco \, cannot penetrate through envelopes
because of large optical depths, but its isotopologues
can be both sufficiently abundant
and optically thin enough to probe central regions close to the protostar. 

In this paper we examine current results for protostellar luminosities as a function of the kinematic masses. 

We find that the luminosities of many objects as a function
of their mass are consistent with 
the protostellar photospheric
radiation being comparable to or larger
than that arising from accretion. 
The implication is that many protostars with mass measurements are currently accreting too slowly for their masses to significantly increase during their expected
lifetimes. There is a hint of possible bimodal distribution, with a fraction of the protostars exhibiting high accretion luminosities and thus high mass addition rates. 
The full implications of these results will await further mass measurements and a careful analysis of observational biases.

\section{Estimated stellar luminosity-mass relations}

The total system luminosities are the sum of stellar photospheric and accretion luminosities. 
As the protostellar emission is not directly
detected but must be inferred from the bolometric
luminosity of the dusty envelope, to make progress
one must make an estimate for
the protostellar photospheric luminosity given its mass.
Furthermore, to make an estimate of the
accretion rate, a further estimate of the protostellar radius as a function of mass is needed, such that
\begin{equation}
L_{bol} = L_* + L_{acc} = L_* + 0.8 G M_* \mdot /R_*\,
\label{eq:lums}
\end{equation}
\citep{gullbring98}.
Here we are assuming that the accretion flow is
not spherical but is rather arising from a disk
or magnetosphere so that there is a distinct protostellar photospheric contribution $L_*$.

\begin{figure}
    \centering
    \includegraphics[width=0.49\textwidth]{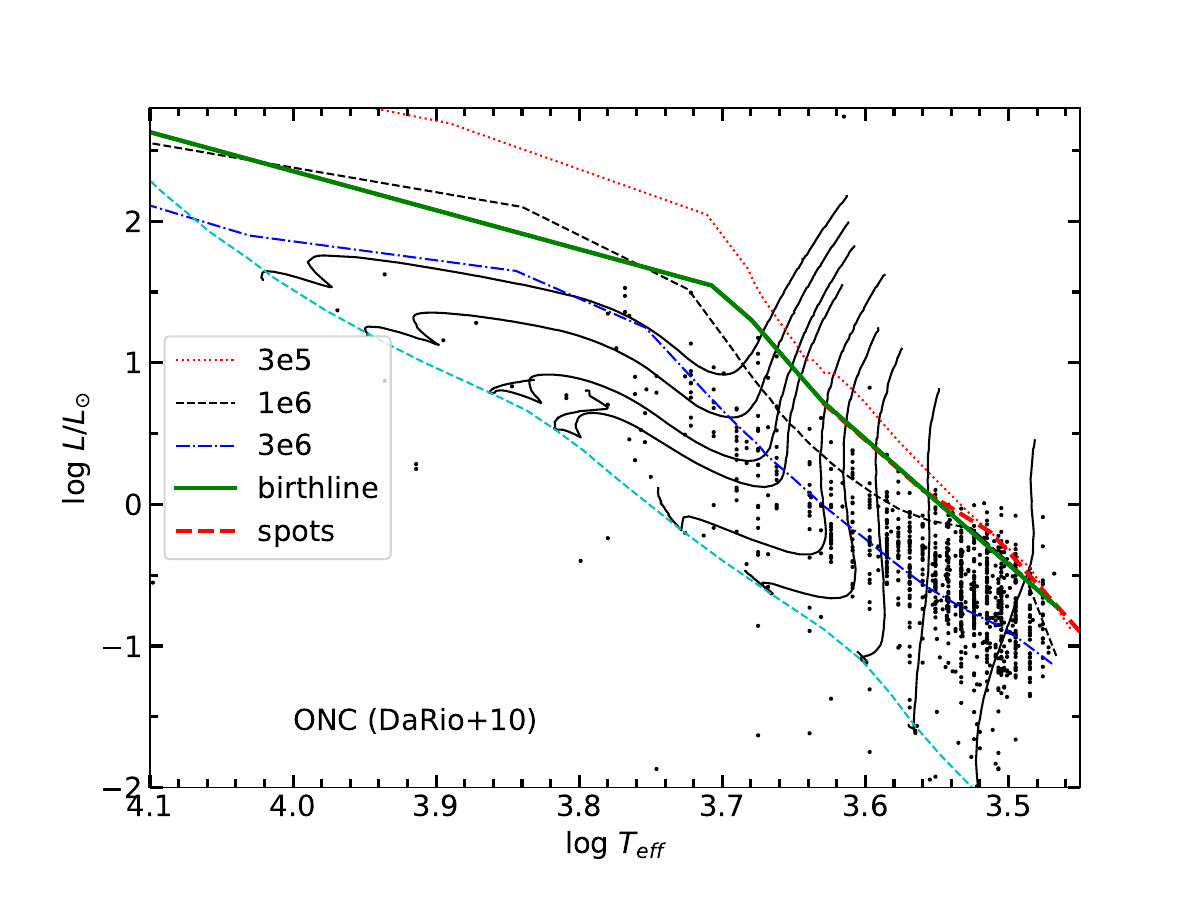}
    \caption{HR diagram of the Orion Nebula
    Cluster stars (points), with our adopted birthline shown as the solid green jagged line. The evolutionary tracks and isochrones for 0.3, 1, and 3 Myr, along with a zero age main sequence, are taken 
    from \protect \cite{siess00}. An alternative birthline using results
    from evolutionary tracks that include the effects of large
    starspots from \protect \cite{cao22} (see also \protect \cite{somers20})
    is shown as a heavy dashed red curve (see text).}
    \label{fig:oncbirthline}
\end{figure}

For an initial exploration we adopt the following
procedure. We assume that protostars have luminosities as a function of their effective temperatures that place them slightly higher in
the HR diagram than
the "youngest" optically-visible (Class II/III) 
stars. This assumes that the youngest directly
detectable stars have contracted only modestly
after their protostellar envelopes have either
completely fallen in or have been otherwise
dispersed, so that the main accretion phase has
ended.
This is essentially the same as the original idea of protostellar
"birthlines" of \cite{stahler83,stahler88}, except
that rather than being tied to theoretical models
including deuterium fusion we use empirical estimates. 
To construct an empirical birthline we use the observed HR diagram of
the well-populated Orion Nebula Cluster from
\cite{dario10}. The main one we use is shown as the solid green curve in Figure \ref{fig:oncbirthline} drawn by eye to lie above almost all of the stars.\footnote{We considered developing a more quantitative method, but decided not to pursue such given the uncertainties in the measurements, complications due to unresolved binaries, and the likelihood that there is no one specific birthline for all stars (see Discussion).}
This birthline corresponds roughly to the
$\sim 0.3 - 0.4$~Myr isochrones of \cite{siess00},
which seems reasonable given a typical observational estimate of $\sim 0.5$~Myr for 
protostellar lifetimes \citep{evans09}.

\begin{figure}
    \centering
    \includegraphics[width=0.45\textwidth]{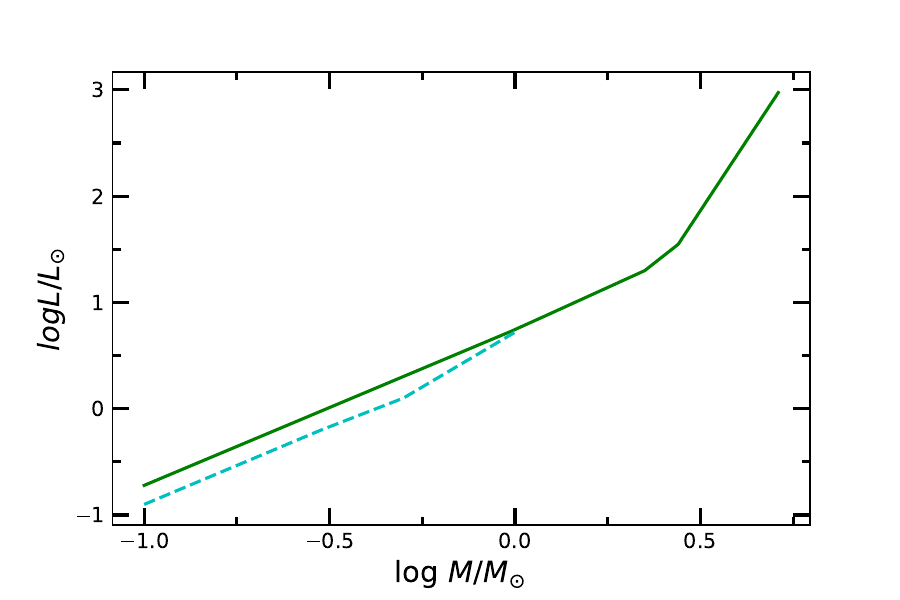}
    \caption{Luminosity-mass relations for the birthlines shown in 
    Figure \ref{fig:oncbirthline}, with the solid curve representing the result for the Siess tracks and the dashed for the SPOTS tracks, as discussed in the text. (At higher masses
    the Siess results are used.)}
    \label{fig:luminosity_mass}
\end{figure}

\begin{figure}
    \centering
    \includegraphics[width=0.45\textwidth]{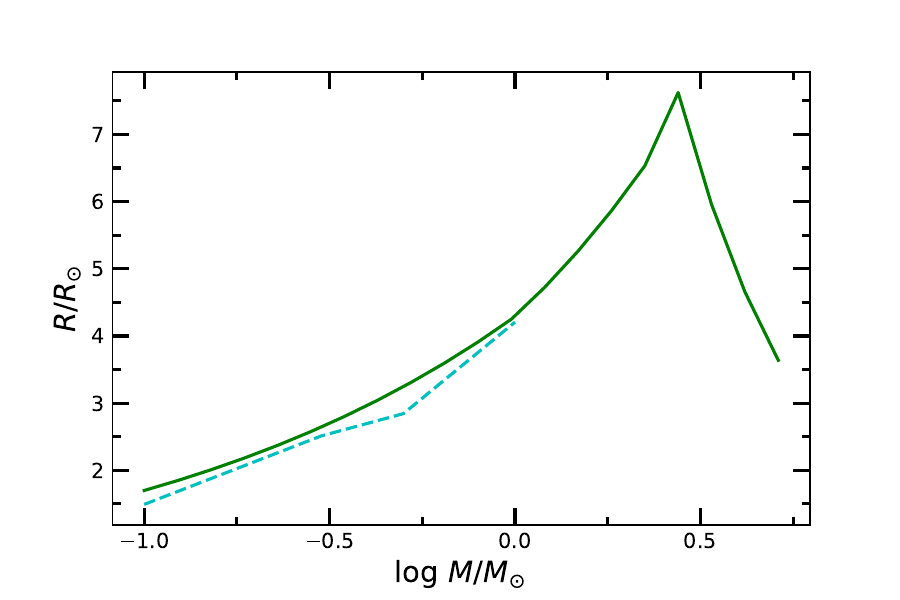}
    \caption{Radius -mass relations for the birthlines shown in 
    Figure \ref{fig:oncbirthline}, with the solid curve representing the result for the Siess tracks and the dashed for the SPOTS tracks.}
    \label{fig:radius_mass}
\end{figure}

To make further progress we need $L_*$ and $R_*$ as a function of
$M_*$, and these must come from theoretical tracks. Our main results 
come from the birthline (solid curve in Figure \ref{fig:oncbirthline})
calibrated in mass from the \cite{siess00} tracks, which
are conventional evolutionary models. However, there is evidence
that the large starspot coverage of young stellar photospheres can
significantly shift evolutionary tracks, particularly for low-mass stars. To address this we examine SPOTS evolutionary models \citep{somers20} to construct an alternative birthline and
luminosity and radius-mass relations. Specifically, we adopted
values of $L_*$, $R_*$ and $M_*$ for the spot coverage fraction $f = 0.34$ models
(as shown for example in Figure 10 of \cite{cao22}, who employed these models to study the $\lambda$ Ori young cluster). We chose an age $\sim 0.3$~Myr
to roughly match the empirical birthline in the
HR diagram.
This SPOTS birthline differs slightly from that adopted for
the Siess tracks (dashed curve in Figure \ref{fig:oncbirthline}).
 While the SPOTS Hayashi tracks significantly change estimated masses from HR diagram positions 
compared with the Siess tracks, due to the sensitivity of the mass to the effective temperature,
Figures \ref{fig:luminosity_mass} and \ref{fig:radius_mass} show modest differences in the
$L_* (M_*)$ and $R_*(M_*)$ ($\sim$
0.2 dex at most in luminosity) that are not significant for our purposes, especially
considering the uncertainties in
the mass and luminosity measurements.

\begin{table}
\tabcolsep6.5pt
\caption{Protostar properties}
\label{tab1}
\begin{center}
\begin{tabular}{@{}l|c|c|c|c|c|l@{}}
\hline
Source & M$_u$ & M$_l$ & Mass & T$_{\mathrm bol}$ & L$_{\mathrm bol}$ & Ref. \\
       &  ($\msun$)  & ($\msun$)    & ($\msun$) & (K)            & (L$_{\odot}$)  &           \\
\hline
HOPS-370   &  ...  &     ...  &   2.5  &   71   &   314  & 1\\
RCrA IRS7B & 3.21  &     2.09  &  2.65 &   88   &   5.1  & 2 \\
L1527 IRS  & 0.49  &     0.32  & 0.41 &   41   &   1.3  & 3 \\
L1489 IRS  & 1.91  &    1.5  & 1.7   &  213   &   4.5  & 4 \\
IRAS 04302+2247  & 1.65 & 1.23 &  1.44 &   88   &  5  & 5\\
CB68       & 0.158   &    0.137    & 0.15  &   50    &   0.89  & 6\\
Ced110 IRS4 & 1.45  &   1.21   & 1.33  &   68   &   1.0 & 7\\
IRAS 16253-2429   & 0.17 &  0.12 & 0.15    &   42   &  0.16  & 8\\ 
RCrA IRS5N & 0.4  &     0.18 &  0.29   &   59   &  1.4  & 9 \\
Oph IRS43-A  &  ... & ...     & 1.0    &  193   &  4.1  & 10\\
Oph IRS43-B  &  ... & ...     & 1.0    &  193   &  4.1  & 10\\
Oph IRS63  &  0.66 & 0.33     & 0.5    &  348   &  1.3   & 11 \\
TMC1A      &  ...   & ...     & 0.68   &  183   &  2.7   & 12 \\
B335       & ...  &   ...     & 0.12  &   41   &  1.4  & 13 \\ 
Lupus 3 MMS &   ... & ... &     0.3    &   39   &  0.41 & 14 \\ 
VLA1623$^{\rm a}$    &   ... & ... &      0.2    &   50   &  1.1  & 15\\
L1455 IRS 1 &  ... & ... &     0.28    &   59  &  3.6  &  16 \\
L1448IRS3B$^{\rm a}$  &  ... &    ...  & 1.19    &   61  &  5.8  & 17 \\
L1448IRS3A  &  ... &    ...  & 1.51    &   47  &  5.8  & 17 \\
TMC1$^{\rm a}$        &  ... &    ...  & 0.54    &   101  &  0.9  & 18\\
L1551NE$^{\rm a}$     &  ... &    ...  &  0.8    &   91  &  4.2  & 19 \\
Elias29     &  ... &    ...  &  2.5    &   350  & 14.1  & 20 \\
HH212 MMS    &  ... &    ...  &  0.2    &  53 & 14    & 21 \\
HH111 MMS    &  ... &    ...  &  1.8    &  78   & 23  & 22 \\
L1551 IRS5$^{\rm a}$   &  ... &    ...  &  0.5    &  94   & 22.1  & 23 \\
IRAS16293-2422-Aa & ... &  ... & 0.9   &  54  &  36   & 24 \\ 
IRAS16293-2422-Ab & ... & ...  & 0.8   &  54  &  36   & 24 \\ 
HOPS-361-A$^{\rm a}$        & 5.46 &  4.40 & 4.93   &  69  &   368 & 25 \\ 
HOPS-361-C$^{\rm a}$        & 1.57 & 1.38  & 1.48   &  69  &  85   & 25 \\ 
GSS30 IRS3          & 0.44   & 0.26 & 0.35 & 50 & 1.7 & 26  \\
IRAS 04169+2702 & ... & ... & 1.0 & 163 & 1.5 & 27 \\
\hline
\end{tabular}
\end{center}

Data used in Figures \ref{fig:luminosity_mass} and \ref{fig:lm_ev}, with
M$_u$ and $M_l$ being upper and lower mass estimates from different
fitting, and Mass being the final average. The values are mostly taken from
\cite{tobin24}, with the following exceptions:
For IRAS 16253-2429 we adopt a mass range of 0.12-0.17 $\msun$, average $0.15 \msun$ from \cite{aso23}. We adopted $0.68 \msun$ and $2.7 \lsun$ for TMC1A from \cite{aso15}. For B335 we used 0.12 $\msun$ and $1.4 \lsun$ from \cite{yen15a}. We added results for GSS30 IRS3 from \cite{santamaria24} and IRAS 04169+2702 from \cite{han23} and \cite{furlan08}. We did not include IRAS 15398-3359. 

$^{\rm a}$Circumbinary mass measurement.

References: 1, \cite{tobin20b}; 2,\cite{ohashi23}; 3, \cite{vanthoff23}, \cite{furlan08}; 4, \cite{yamato23},\cite{furlan08};
5, \cite{lin23},\cite{villenave24}; 6, \citep{kido23}; 7, \citep{sai23};
8, \citep{aso23}; 9, \citep{sharma23}; 10, \citep{brinch16};
11, \citep{flores23}; 12, \citep{ohashi23}; 13, \citep{yen15a}; 
14, \citep{yen17}; 15, \citep{murillo13}; 16, \citep{chou16};
1k, \citep{reynolds21}; 18, \citep{harsono14};
19, \citep{takakuwa12}; 20 \citep{lommen08}; 21 \citep{lee17};
22 \citep{tobin20a}; 23 \citep{chou14}; 24, \citep{maureira20}; 
25, \citep{cheng22}; 26, \citep{santamaria24};
27 \citep{han23} \\
\end{table}

\begin{figure}
    \centering
    \includegraphics[width=0.49\textwidth]{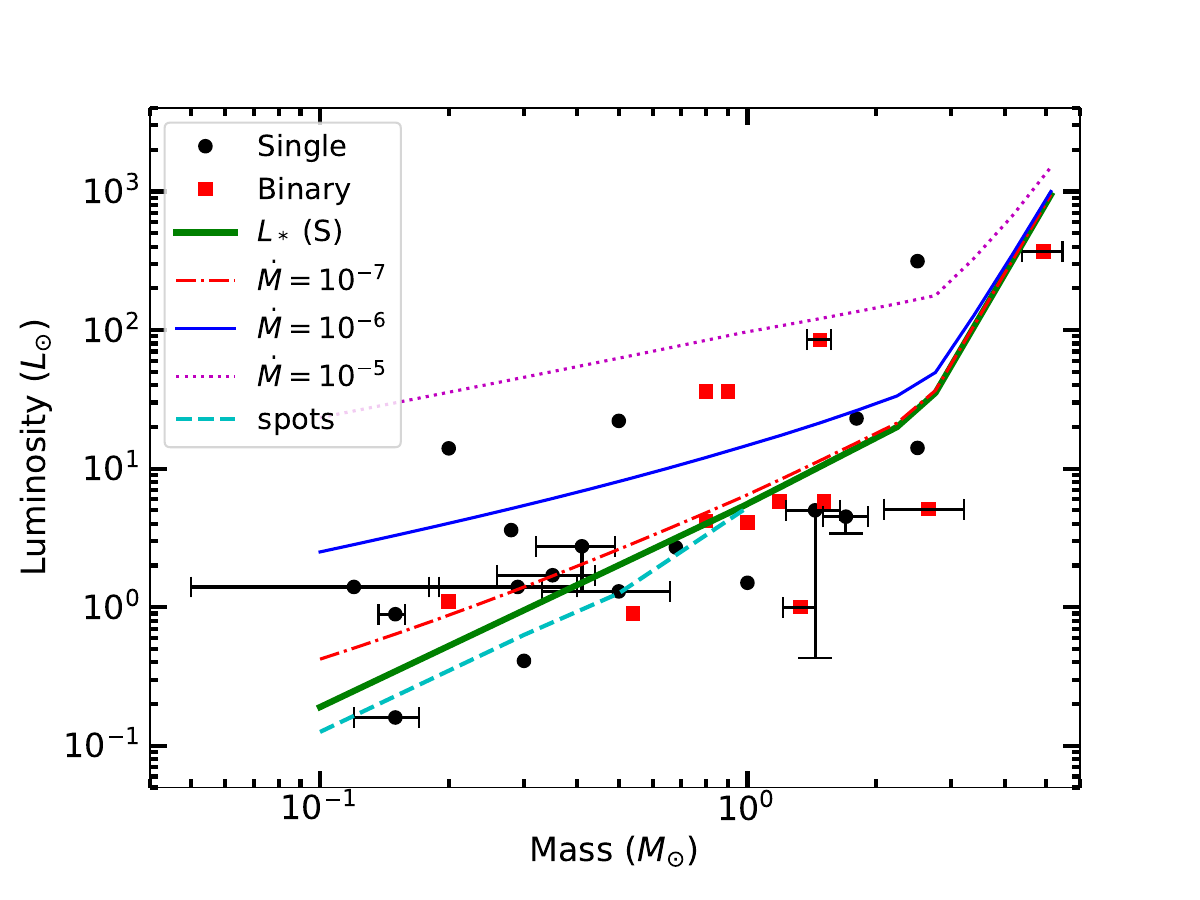}
    \caption{Comparison of birthline models to observed protostellar masses and bolometric luminosities. The green solid curve is for the birthline using Siess tracks, and the dashed cyan curve is for the SPOTS tracks and birthline. The red dot-dashed, the
    solid blue, and the dotted magenta curves add the accretion luminosity
    to the photospheric values for $10^{-7} \msunyr$, $10^{-6} \msunyr$ and $10^{-5} \msunyr$ respectively, adopting the
    Siess birthline and $R_*$ vs. $M_*$ relation. When
    available, error estimates for the protostellar masses
    are shown as horizontal ranges. The vertical bars
    are for cases in which the observed bolometric
    luminosity is lower than a model result that takes
    into account the non-isotropic radiation of the 
    star and envelope (see text).}
    \label{fig:proto_lm}
\end{figure}

\section{Results}

The sample of protostar luminosity and mass measurements displayed in Figure \ref{fig:proto_lm} come from Table 1, mostly taken from \citep[][and references therein]{tobin24} 
with a few exceptions as noted below.
We show mass error bars
in cases where error estimates or ranges are quoted.
For binary systems we show individual mass measurements where available, but the luminosities
are mostly that of the total system, as there is 
generally no way of separating the contributions of the components. 
For comparison, we show photospheric
birthlines for the Siess and the SPOTS tracks, along with the sum
of photospheric and accretion luminosities for the specified
mass accretion rates, using equation \ref{eq:lums} and the Siess luminosity-mass and radius-mass relations (Figures \ref{fig:luminosity_mass} and \ref{fig:radius_mass}).

While the luminosities of some protostars lie below the birthlines, this may be due to
a systematic bias arising from
the assumption that their envelopes absorb all the light
from the central regions and reradiate this energy isotropically.
This is generally not the case as envelopes are often flattened
or otherwise asymmetric, with bipolar outflow cavities along
which radiation escapes more readily, resulting in sources
appearing underluminous viewed in other directions. This is
particularly true for disk systems viewed edge-on.
An extreme example of this is 
IRAS 04302+2247, which is an edge-on disk system
with a tenuous surrounding envelope \citep{villenave24}. The nominal observed bolometric luminosity $0.43 \lsun$ \citep{lin23} is much lower than the $1.8 \lsun$ estimated from the SED modeling
of \cite{furlan08}, and especially the $5 \lsun$ adopted in the modeling of \cite{villenave24}; we adopt this last value. L1527 is another
edge-on disk system where the observed
bolometric luminosity is $1.3 \lsun$ while detailed modeling including analysis of the scattered light in the outflow cavities yields $2.75 \lsun$ 
\citep{tobin08}.
For these objects, as well as for L1489IRS (where we adopt $4.5 \lsun$ from \citealt{furlan08}), we display the results along vertical lines, where the lower limit is the ``observed'' bolometric luminosity,
while the data point is the result from the
radiative transfer modeling.


Even when not observed edge-on, estimates of protostellar luminosities may be biased toward lower values. 
For example, radiative transfer solutions for the \cite{terebey84}
envelope models with outflow cavities
generally
predict apparent luminosities factors of $\sim 1.5-2$ smaller than true values 
\citep{whitney03}. (Conversely, protostars will appear
overluminous when observed down outflow cavities (face-on), but these objects may not be recognized as Class 0/I systems.)

\section{Discussion}

The most striking result in Figure \ref{fig:proto_lm} is the clustering of protostars 
near the birthlines, suggesting that in many
systems the protostellar photospheric radiation is a significant
if not the main component of the total luminosity.
The adopted birthlines
near the conventional i
$\sim 0.3 - 0.4$ Myr isochrones seem consistent with
typical estimates of protostellar lifetimes \citep[e.g.,][]{evans09},
but apart from a general statement that the isochronal age should not be much longer than the protostellar collapse time
(otherwise the star would contract to lower luminosities),
the actual details of protostar formation are likely to be
considerably different than contraction starting with very large
radii.

Detailed calculations for spherical collapse, starting with \cite{larson69},
yield
birthlines that in some cases lie well
above those used here 
\citep[depending upon infall/accretion rates;][]{stahler88,palla91,palla92}.
However, it is likely that a significant fraction of the stellar mass is not
accreted spherically but through an accretion disk, whose
geometry allows easier escape of radiation and thus generally
lower stellar radii at the end of major accretion 
\citep{hartmann97,baraffe09,baraffe12,baraffe17}. In addition, the importance of deuterium fusion depends upon initial conditions \citep{baraffe12}. As it is unlikely that all systems form with the same angular momenta, initial densities, etc. it would not be
surprising if protostars of the same mass and age have somewhat different initial conditions and thus populate
a ``birth region'' rather than a single
line. But even if our birthlines somewhat overestimate photospheric contributions, resulting in higher accretion luminosities, the mass accretion rates would still be low ($\sim 10^{-7} \msunyr$).

A modest fraction of the protostar sample exhibit luminosities as much as an order of magnitude larger than the near-birthline objects, and are most likely dominated by accretion.
The large separation between the bright and faint protostars is suggestive of a
bimodal distribution. 
However, this is not
definitive given small number statistics; in addition,
the sample biases are not well understood.

\begin{figure}
\centering
\includegraphics[width=0.49\textwidth]{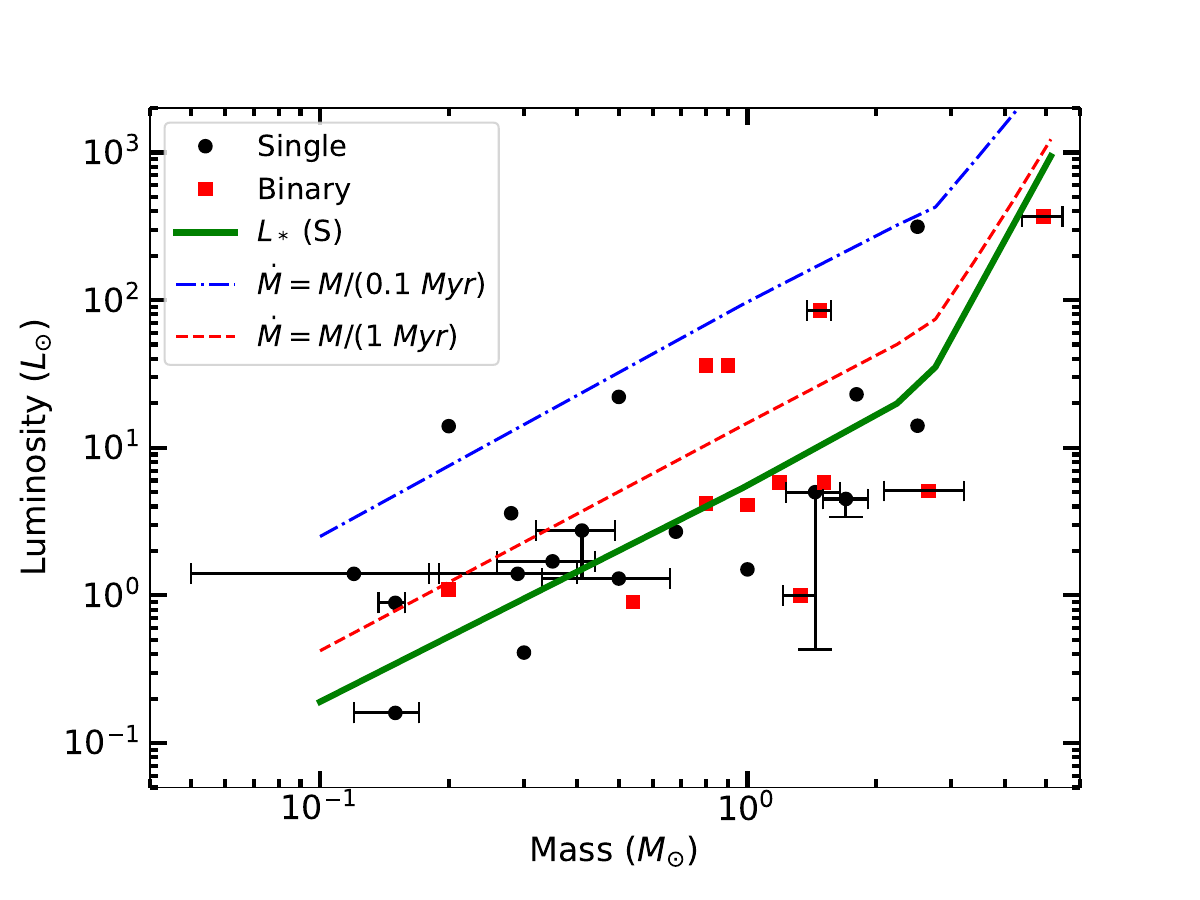}
\caption{Same data and Siess photospheric luminosity curve as in Figure \ref{fig:luminosity_mass}, but now with added accretion luminosities for accretion rate equal to the mass divided by 1 Myr (lower dashed line)
and 0.1 Myr (upper dot-dashed line.)}
\label{fig:lm_ev}
\end{figure}

\begin{figure}
\centering
\includegraphics[width=0.49\textwidth]{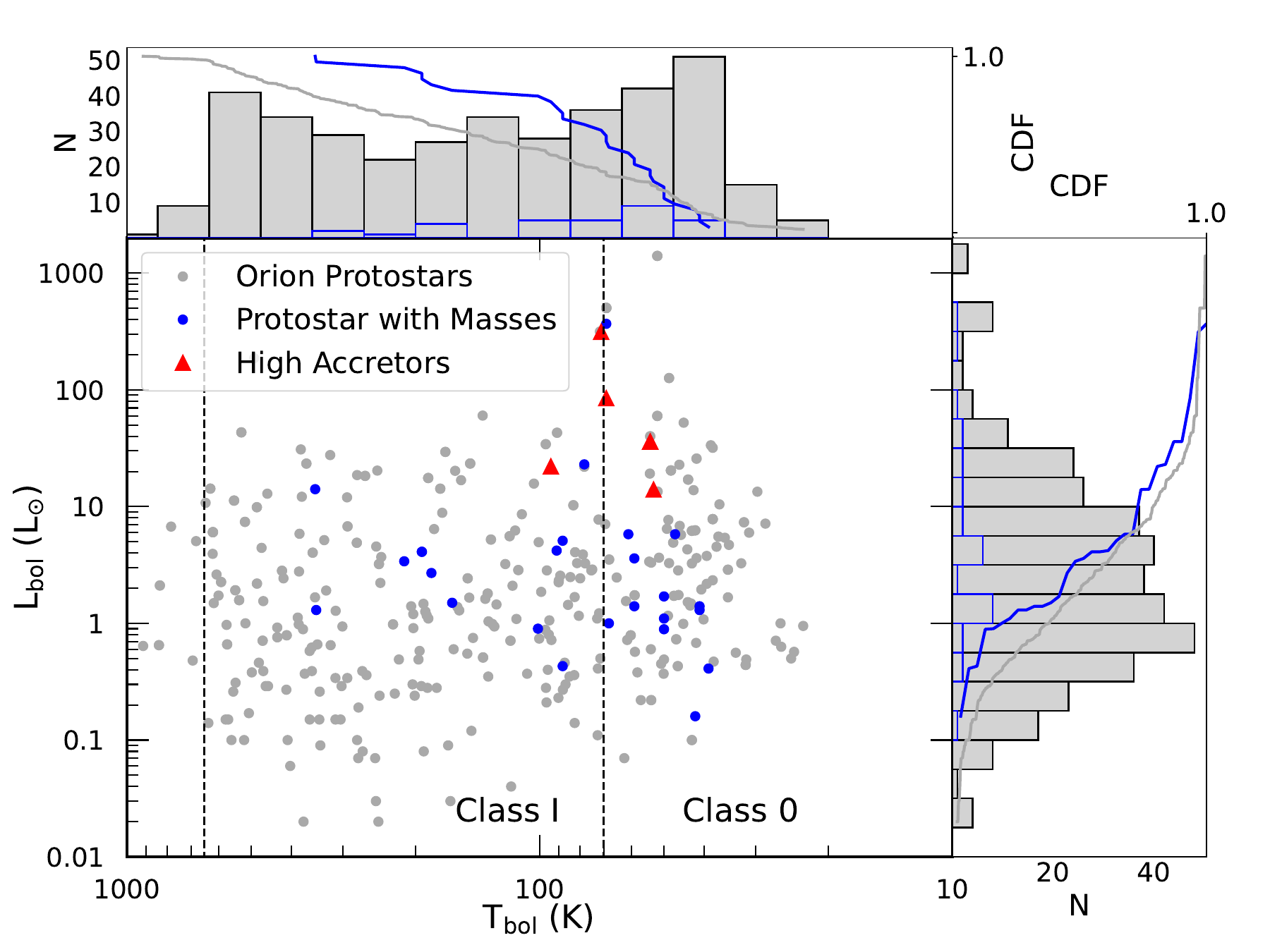}
\caption{Observed luminosities of protostars vs. \tbol\ for the Herschel Orion Protostar Survey (HOPS; \protect \cite{furlan16}, shown as gray points, with the protostars with masses highlighed with large blue dots. The rapid accretors are shon as red triangles; 
IRAS 16293Aa and Ab are plotted as one symbol
as they share
the same \lbol, \tbol, and disk flux density and radius.
The upper and side panels show binned data along with cumulative distributions. The upper panel shows that the distribution of protostars with masses is more weighted toward \protect \tbol and Class 0 objects, and the right panel shows that the luminosity distribution of protostars with mass is also shifted slightly toward higher luminosities.}
\label{fig:lbol_tbol}
\end{figure}

To emphasize the evolutionary implications of the
current mass-luminosity relation,
in  Figure \ref{fig:lm_ev} we show
the predicted luminosities for
accretion rates that would double
the mass on timescales of 1 Myr and 0.1 Myr. With typical estimated protostellar lifetimes
of 0.5 Myr, it appears that
many objects in the current sample are not accreting at rates that would
significantly change their masses. 
A subset of higher luminosity protostars
are consistent with
evolutionarily-important accretion.
One possibility is that the lower luminosity protostars are ``old'', and have mostly finished accumulating
their masses, while the bright protostars are younger.
However, inspection of the distribution of bolometric temperatures \citep[which are generally taken to be a guide to evolutionary state;][]{myers93} vs. luminosity shows that the sample with mass measurements tends to favor earlier evolutionary phases (Figure \ref{fig:lbol_tbol}). Thus, it is not obvious that the current sample is biased against the
youngest objects.

Alternatively, if accretion outbursts dominate the mass addition,
\citep[e.g.,][]{kenyon90}, then
the spread in observed luminosities could simply reflect burst behavior.
With the small number statistics of the current sample of protostars, and probable observational selection effects, it is not possible to distinguish whether evolutionary effects (ages) or bursts dominate, or even whether both contribute.

\begin{figure}
    \centering \includegraphics[width=0.49\textwidth]{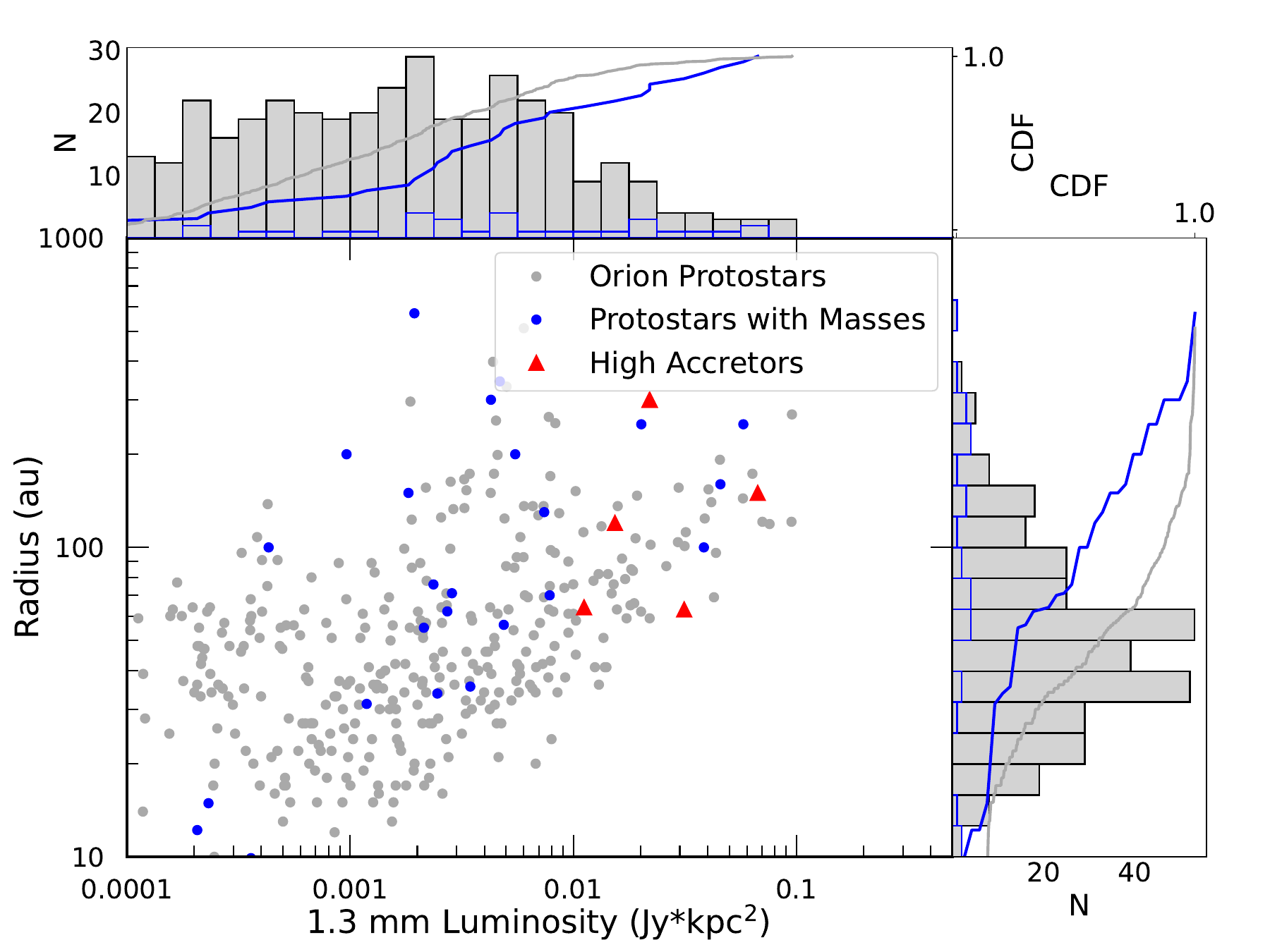}
    \caption{Disk radii vs. mm-wave luminosity (proportional to dust mass) for the HOPS sample and the subsample with mass measurements, with symbols, binned data, and cumulative distributions as in Figure \ref{fig:lbol_tbol}. There is a clear bias toward larger disks. The Orion data from \citet{tobin20a} has its mm-luminosities scales for 0.89~mm to 1.3~mm assuming an average spectral index of 2.5. There appears to also be a bias toward higher luminosity disks in the protostars with measured masses, but this is less certain given the scaling of Orion protostars to 1.3~mm. }
    \label{fig:flux_r}
\end{figure}

Figure \ref{fig:flux_r} shows the 1.3mm continuum disk radii as a function of the mm flux. Unsurprisingly, there is a bias toward
larger disks in the sample with masses, as it is easier to resolve the Keplerian rotation.
If disks with larger radii have more mass, then it might indicate that the current results are weighted toward more massive protostars. In any case
larger samples with better control of selection criteria are needed to
make further progress in understanding protostellar accretion and the distribution of protostar masses.



\section*{Acknowledgements}

LH was supported in part by
NASA Emerging Worlds grant 80NSSC24K1285.
J.J.T. acknowledges support from NASA XRP 80NSSC22K1159. 
The National Radio Astronomy Observatory and Green Bank Observatory are facilities of the U.S. National Science Foundation operated under cooperative agreement by Associated Universities, Inc.
Software: Matplotlib \citep{Matplotlib}; \texttt{python} \citep{python}.
Software citation information aggregated using \texttt{\href{https://www.tomwagg.com/software-citation-station/}{The Software Citation Station}} \citep{software-citation-station-paper, software-citation-station-zenodo}.
This research has made use of NASA’s
Astrophysics Data System Bibliographic Services.
\section*{Data availability}
The observational data upon which this paper is based have been published
elsewhere and are summarized in
Table 1. Birthlines will be made available by the authors upon reasonable request.

\section{Other differences from Tobin and Sheehan}
For IRAS 16253-2429 we adopt a mass range of 0.12-0.17 $\msun$, average $0.15 \msun$ from \cite{aso23}. We adopted $0.68 \msun$ and $2.7 \lsun$ for TMC1A from \cite{aso15}. For B335 we adopted a range 0.05-0.19 $\msun$ from \cite{yen15a}. We added results for GSS30 IRS3 from \cite{santamaria24}. We added results for IRAS 04169+2702 from \cite{han23} and \cite{furlan08}. We did not include IRAS 15398-3359.

\bibliographystyle{mnras}
\bibliography{refs1}

\label{lastpage}

\end{document}